\documentclass[10pt,journal]{IEEEtran}
\usepackage{amsmath,amsfonts}
\usepackage{algorithmic}
\usepackage{array}
\usepackage[caption=false,font=normalsize,labelfont=sf,textfont=sf]{subfig}
\usepackage{textcomp}
\usepackage{stfloats}
\usepackage{url}
\usepackage{bm}
\usepackage{cuted}
\usepackage{longtable}
\usepackage{verbatim}
\usepackage{graphicx}
\usepackage{amsmath}
\usepackage{lineno,hyperref}
\usepackage{multirow}
\usepackage{makecell}
\usepackage{pdflscape}
\usepackage{rotating}
\usepackage{booktabs} 
\usepackage{tabularx} 
\usepackage{caption}  
\usepackage{array}      
\usepackage{booktabs}   
\usepackage{cite}       

\usepackage{stfloats}
\def\BibTeX{{\rm B\kern-.05em{\sc i\kern-.025em b}\kern-.08em
    T\kern-.1667em\lower.7ex\hbox{E}\kern-.125emX}}
\usepackage{balance}
\begin{document}
\title{Foundation Models for Cross-Domain EEG Analysis Application: A Survey}
\author{Hongqi Li~\IEEEmembership{Member,~IEEE}, Yitong Chen, Yujuan Wang, Weihang Ni, Haodong Zhang* 
\thanks{Manuscript received Aug 12, 2025. This work was supported in part by the Natural Science Basic Research Program of Shaanxi Province under Grant 2024JC-YBQN-0659, in part by the Student Innovation Foundation of Northwestern Polytechnical University Taicang Innovation Harbor under Grant TCCX250103, and in part sponsored by the Practice and Innovation Funds for Graduate Students of Northwestern Polytechnical University under Grant PF2025013. (\textit{Corresponding author: Haodong Zhang}.)

H. Li, Y. Chen, Y. Wang, W. Ni, and H. Zhang are with the School of Software, Northwestern Polytechnical University, Xi’an 710729, China (e-mail: lihongqi@nwpu.edu.cn, chenyt@mail.nwpu.edu.cn, wangyujuan12@mail.nwpu.edu.cn,  niweihang@mail.nwpu.edu.cn, zhang\_haodong@\\mail.nwpu.edu.cn)
}}

\markboth{Journal of \LaTeX\ Class Files,~Vol.~18, No.~9, September~2020}%
{How to Use the IEEEtran \LaTeX \ Templates}

\maketitle
\begin{abstract}
Electroencephalography (EEG) analysis stands at the forefront of neuroscience and artificial intelligence research, where foundation models are reshaping the traditional EEG analysis paradigm by leveraging their powerful representational capacity and cross-modal generalization. However, the rapid proliferation of these techniques has led to a fragmented research landscape, characterized by diverse model roles, inconsistent architectures, and a lack of systematic categorization. To bridge this gap, this study presents the first comprehensive modality-oriented taxonomy for foundation models in EEG analysis, systematically organizing research advances based on output modalities of the native EEG decoding, EEG-text, EEG-vision, EEG-audio, and broader multimodal frameworks. We rigorously analyze each category’s research ideas, theoretical foundations, and architectural innovations, while highlighting open challenges such as model interpretability, cross-domain generalization, and real-world applicability in EEG-based systems. By unifying this dispersed field, our work not only provides a reference framework for future methodology development but accelerates the translation of EEG foundation models into scalable, interpretable, and online actionable solutions. 
\end{abstract}


\captionsetup{
  labelsep=space,     
}
\section{Introduction}
\label{chap:1}
\IEEEPARstart{E}{lectroencephalography} (EEG) is a non-invasive technique for recording electrical activity in the brain. It employs scalp electrode arrays to capture multi-channel time-series voltage signals at various sampling frequencies, typically exhibiting low signal-to-noise ratios, non-stationarity, and substantial inter-subject variability. Consequently, developing a robust decoding model with strong cross-subject generalization capabilities is of paramount importance for identifying user intentions and the further brain-machine interactions \cite{1,2}. 

Traditional machine learning approaches \cite{3} rely on manually designed features, limiting their adaptability to complex spatiotemporal patterns. In contrast, end-to-end deep learning (DL) architectures, such as convolutional neural networks (CNNs), recurrent neural networks (RNNs), Transformers, and graph neural networks (GNNs), have demonstrating superior performance by autonomously learning hierarchical representations \cite{4,5}. Nevertheless, their efficacy often necessitate large amounts of high-quality labeled data, which are scarce for EEG due to acquisition costs and inter-individual variability.

To mitigate data scarcity, researchers are increasingly turning to the transfer learning and pre-training strategies. In recent years, the paradigm of foundation models \cite{6} has emerged, wherein neural networks are initially pre-trained on large-scale text, visual, or audio datasets and subsequently adapted to downstream tasks by minimal fine-tuning or direct extraction of intermediate representations. Notable examples include GPT \cite{7} and LLaMA \cite{8} in the realm of language processing, ViT \cite{9} and Swin Transformer \cite{10} in computer vision, Wav2Vec \cite{11} and Whisper \cite{12} in audio processing, CLIP \cite{13} and Flamingo \cite{14} as vision-language models, and Gemini \cite{15} PaLM‑E \cite{16} as fully multimodal architectures, etc. Owing to the powerful representation learning capabilities, zero- or few-shot transfer abilities, and effective cross-modal alignment mechanisms, these mentioned models have exhibited remarkable performance across their respective fields.

\makeatletter
\setlength{\dbltextfloatsep}{3pt} 
\setlength{\dblfloatsep}{3pt}     
\setlength{\@dblfptop}{0pt}       
\setlength{\@dblfpsep}{6pt}
\setlength{\@dblfpbot}{0pt}
\makeatother


\begin{figure*}[t]
  \centering
  \includegraphics[
    pagebox=cropbox, 
    clip,
    trim = 16mm 155mm 16mm 30mm, 
    width=\textwidth 
  ]{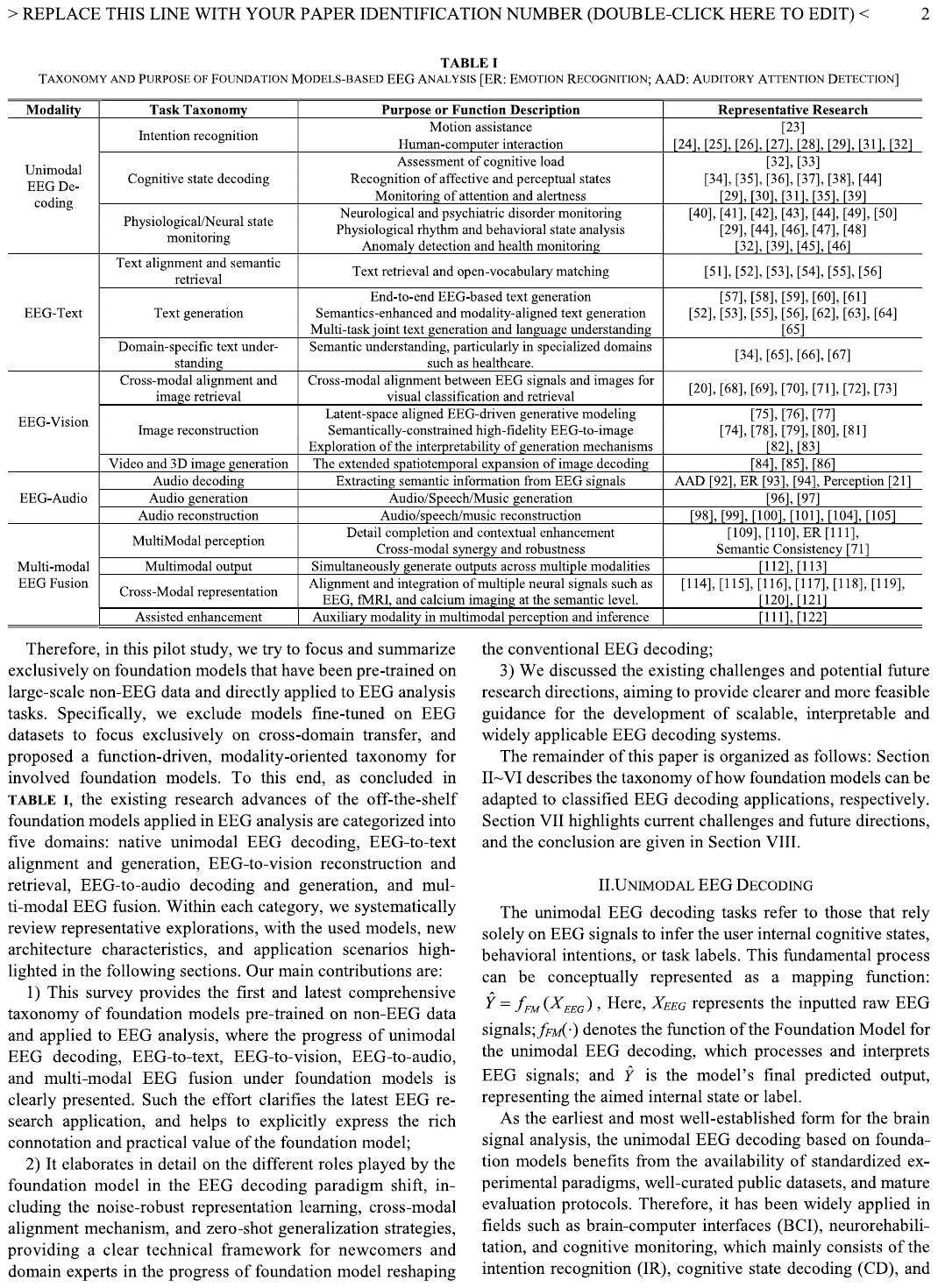}
\end{figure*}

Inspired by these successful cases, researchers began to explore the feasibility of foundation model-based EEG analysis \cite{17}. Preliminary studies suggest that the employment of these pre-trained models can significantly enhance the decoding effects of EEG-to-text \cite{18,19}, EEG-to-image \cite{20}, and EEG-to-audio tasks \cite{21}. As a result, a large number of related studies are being produced in an explosive manner, and task has evolved from traditional closed-set classification toward open-ended generation and cross-modal semantic alignment. But, at the same time, serious fragmentation problems in model roles, architectures, and application domains have also emerged. One primary factor lies in the black-box nature of DL architectures, which allows for extensive customization in signal preprocessing, feature extraction, and classification modules. Meanwhile, foundation models have attempted different roles during the process of task paradigm transition. In particular, apart from serving as feature extractors, large-scale pretrained models can also be used as alignment tools (bridging EEG with textual, visual, or auditory representations via contrastive learning) and as generative backbones (enabling neural signal-driven content synthesis). However, to the best of our knowledge, the rapid adoption of foundation models in EEG analysis has outpaced existing reviews \cite{5,22}, which predominantly focus on the models trained by existing EEG datasets. To date, the field still lacks a specialized summary of the increasingly wide-spread application of non-EEG pre-trained foundation models, especially in light of the explosive research in the nearest three years, to clarify the transferability of techniques across domains.

Therefore, in this pilot study, we try to focus and summarize exclusively on foundation models that have been pre-trained on large-scale non-EEG data and directly applied to EEG analysis tasks. Specifically, we exclude models fine-tuned on EEG datasets to focus exclusively on cross-domain transfer, and proposed a function-driven, modality-oriented taxonomy for involved foundation models. To this end, as concluded in TABLE I, the existing research advances of the off-the-shelf foundation models applied in EEG analysis are categorized into five domains: native unimodal EEG decoding, EEG-to-text alignment and generation, EEG-to-vision reconstruction and retrieval, EEG-to-audio decoding and generation, and multi-modal EEG fusion. Within each category, we systematically review representative explorations, with the used models, new architecture characteristics, and application scenarios highlighted in the following sections. Our main contributions are:

1) This survey provides the first and latest comprehensive taxonomy of foundation models pre-trained on non-EEG data and applied to EEG analysis, where the progress of unimodal EEG decoding, EEG-to-text, EEG-to-vision, EEG-to-audio, and multi-modal EEG fusion under foundation models is clearly presented. Such the effort clarifies the latest EEG research application, and helps to explicitly express the rich connotation and practical value of the foundation model;

2) It elaborates in detail on the different roles played by the foundation model in the EEG decoding paradigm shift, including the noise-robust representation learning, cross-modal alignment mechanism, and zero-shot generalization strategies, providing a clear technical framework for newcomers and domain experts in the progress of foundation model reshaping 
the conventional EEG decoding;

3) We discussed the existing challenges and potential future research directions, aiming to provide clearer and more feasible guidance for the development of scalable, interpretable and widely applicable EEG decoding systems.
The remainder of this paper is organized as follows: Section II~VI describes the taxonomy of how foundation models can be adapted to classified EEG decoding applications, respectively. Section VII highlights current challenges and future directions, and the conclusion are given in Section VIII.

\section{Unimodal EEG Decoding}
\label{chap:2}

The unimodal EEG decoding tasks refer to those that rely solely on EEG signals to infer a user's internal cognitive states, behavioral intentions, or task labels. This fundamental process can be conceptually represented as a mapping function \( \hat{y} = f_{\mathrm{FM}}\left(X_{\mathrm{EEG}}\right) \), where \(X_{\mathrm{EEG}}\) represents the input raw EEG signals, \(f_{\mathrm{FM}}(\cdot)\) denotes the Foundation Model for unimodal EEG decoding, and \(\hat{y}\) is the model’s final predicted output.

As the earliest and most well-established form for the brain signal analysis, the unimodal EEG decoding based on foundation models benefits from the availability of standardized experimental paradigms, well-curated public datasets, and mature evaluation protocols. Therefore, it has been widely applied in fields such as brain-computer interfaces (BCI), neurorehabilitation, and cognitive monitoring, which mainly consists of the intention recognition (IR), cognitive state de-coding (CD), and physiological and neurological state monitoring (PNM), as illustrated in Fig. 1. 

\begin{figure}[htpb]
\centering
\includegraphics[width=8.8cm]{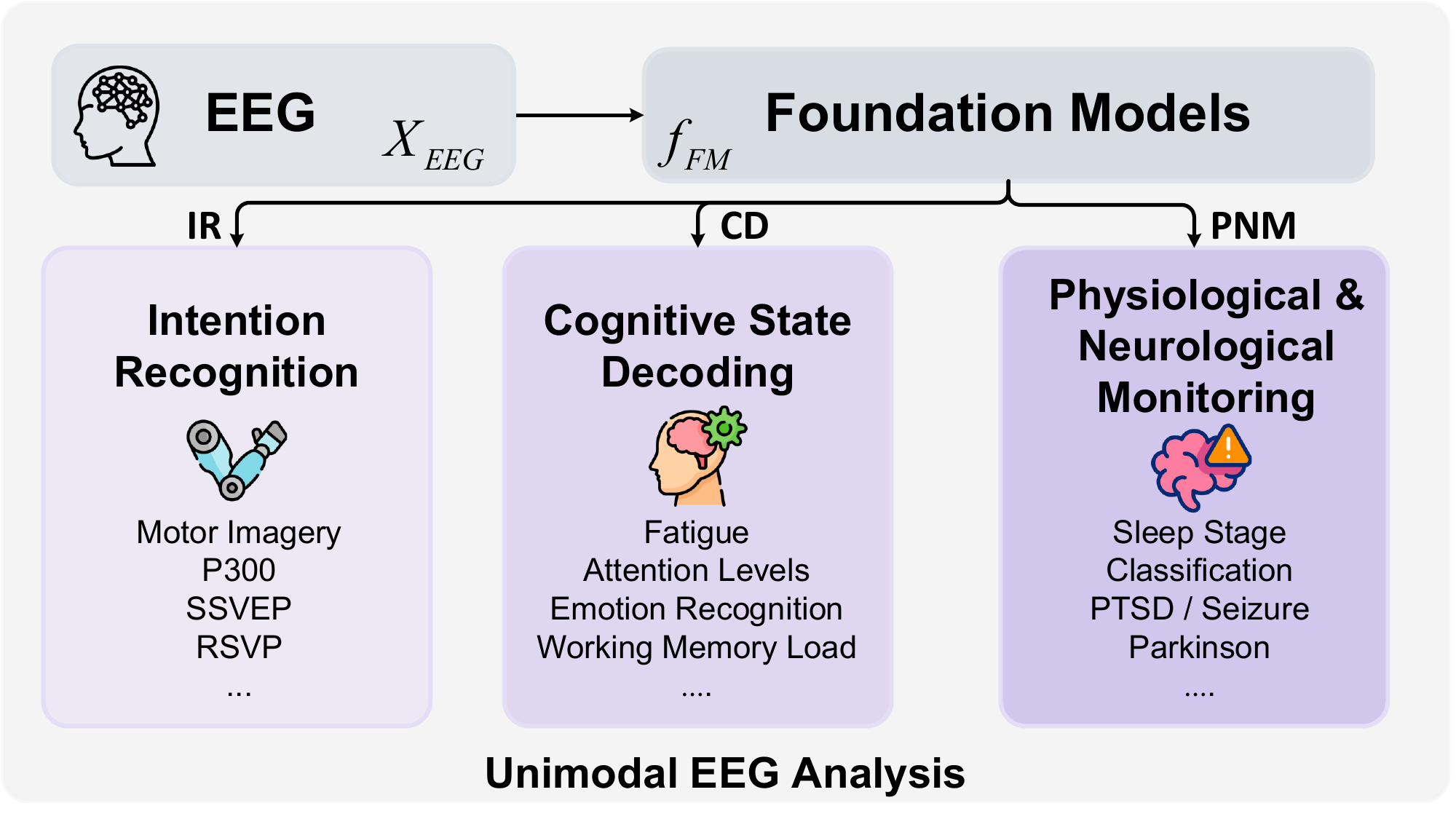}
\captionsetup{font=footnotesize}
\caption{Unimodal EEG decoding tasks based on Foundation Models.}
\label{fig 1}
\end{figure}
\vspace{-0.2cm}

\begin{figure*}[b]
  \centering
  \includegraphics[
    pagebox=cropbox, 
    clip,
    trim = 16mm 32mm 16mm 178mm, 
    width=\textwidth 
  ]{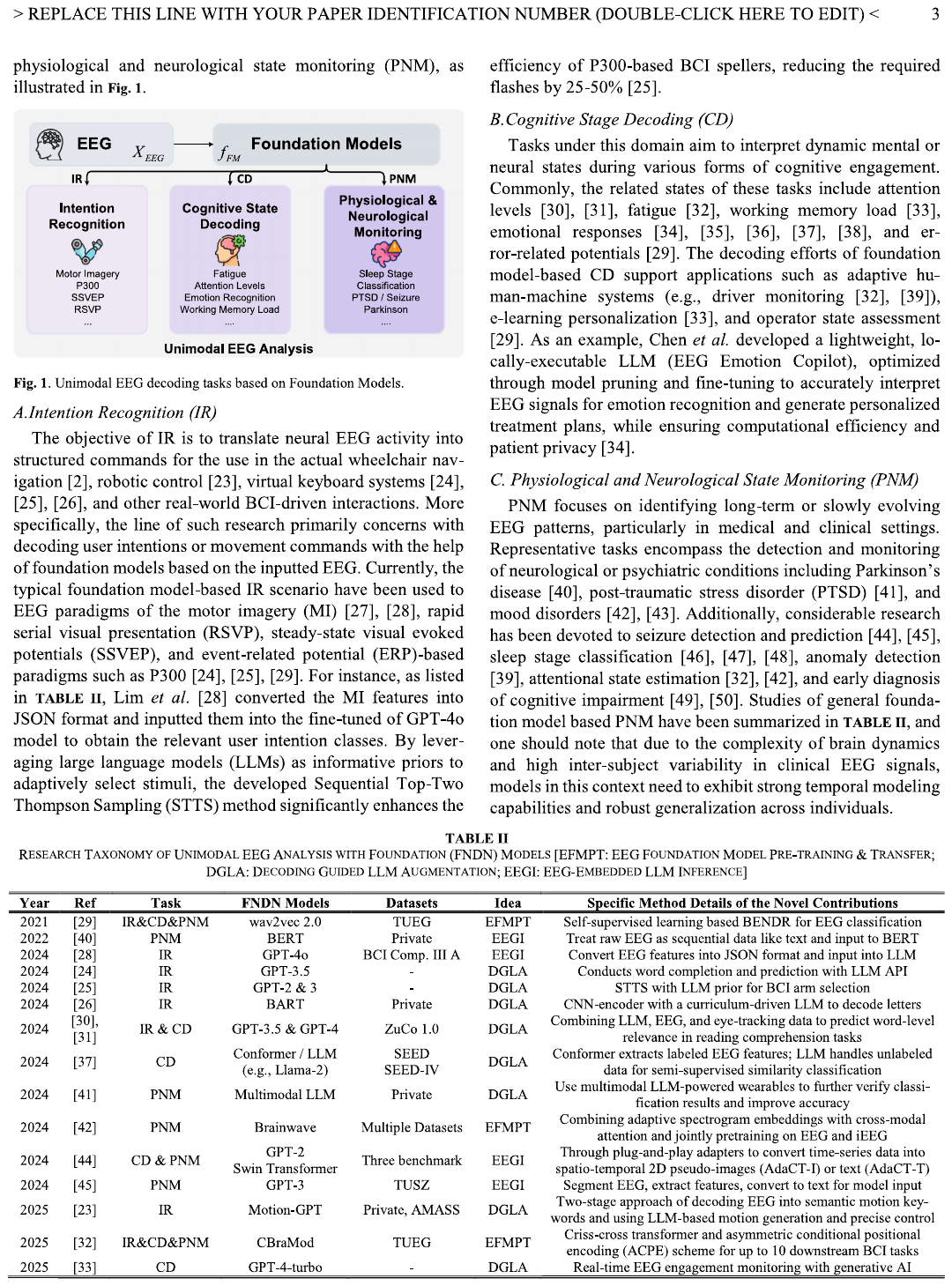}
\end{figure*}

\subsection{Intention Recognition (IR)}
The objective of IR is to translate neural EEG activity into structured commands for the use in the actual wheelchair navigation \cite{2}, robotic control \cite{23}, virtual keyboard systems \cite{24,25,26}, and other real-world BCI-driven interactions. More specifically, the line of such research primarily concerns with decoding user intentions or movement commands with the help of foundation models based on the inputted EEG. Currently, the typical foundation model-based IR scenario have been used to EEG paradigms of the motor imagery (MI) \cite{27,28}, rapid serial visual presentation (RSVP), steady-state visual evoked potentials (SSVEP), and event-related potential (ERP)-based paradigms such as P300 \cite{24,25,29}. For instance, as listed in TABLE II, Lim et al. \cite{28} converted the MI features into JSON format and inputted them into the fine-tuned of GPT-4o model to obtain the relevant user intention classes. By leveraging large language models (LLMs) as informative priors to adaptively select stimuli, the developed Sequential Top-Two Thompson Sampling (STTS) method significantly enhances the efficiency of P300-based BCI spellers, reducing the required flashes by 25-50\% \cite{25}.

\subsection{Cognitive Stage Decoding (CD)}
Tasks under this domain aim to interpret dynamic mental or neural states during various forms of cognitive engagement. Commonly, the related states of these tasks include attention levels \cite{30,31}, fatigue \cite{32}, working memory load \cite{33}, emotional responses \cite{34,35,36,37,38}, and error-related potentials [29]. The decoding efforts of foundation model-based CD support applications such as adaptive human-machine systems (e.g., driver monitoring \cite{32,39}), e-learning personalization \cite{33}, and operator state assessment \cite{29}. As an example, Chen et al. developed a lightweight, locally-executable LLM (EEG Emotion Copilot), optimized through model pruning and fine-tuning to accurately interpret EEG signals for emotion recognition and generate personalized treatment plans, while ensuring computational efficiency and patient privacy \cite{34}. 
\subsection{Physiological and Neurological State Monitoring (PNM)}
PNM focuses on identifying long-term or slowly evolving EEG patterns, particularly in medical and clinical settings. Representative tasks encompass the detection and monitoring of neurological or psychiatric conditions including Parkin-son’s disease \cite{40}, post-traumatic stress disorder (PTSD) \cite{41}, and mood disorders \cite{42,43}. Additionally, considerable research has been devoted to seizure detection and prediction \cite{44,45}, sleep stage classification \cite{46,47,48}, anomaly detection [39], attentional state estimation \cite{32,42}, and early diagnosis of cognitive impairment \cite{49,50}. Studies of general foundation model based PNM have been summarized in TABLE II, and one should note that due to the complexity of brain dynamics and high inter-subject variability in clinical EEG signals, models in this context need to exhibit strong temporal modeling capabilities and robust generalization across individuals. 

\begin{figure*}[t]
  \centering
  \includegraphics[
    pagebox=cropbox, 
    clip,
    trim = 16mm 240mm 16mm 30mm, 
    width=\textwidth 
  ]{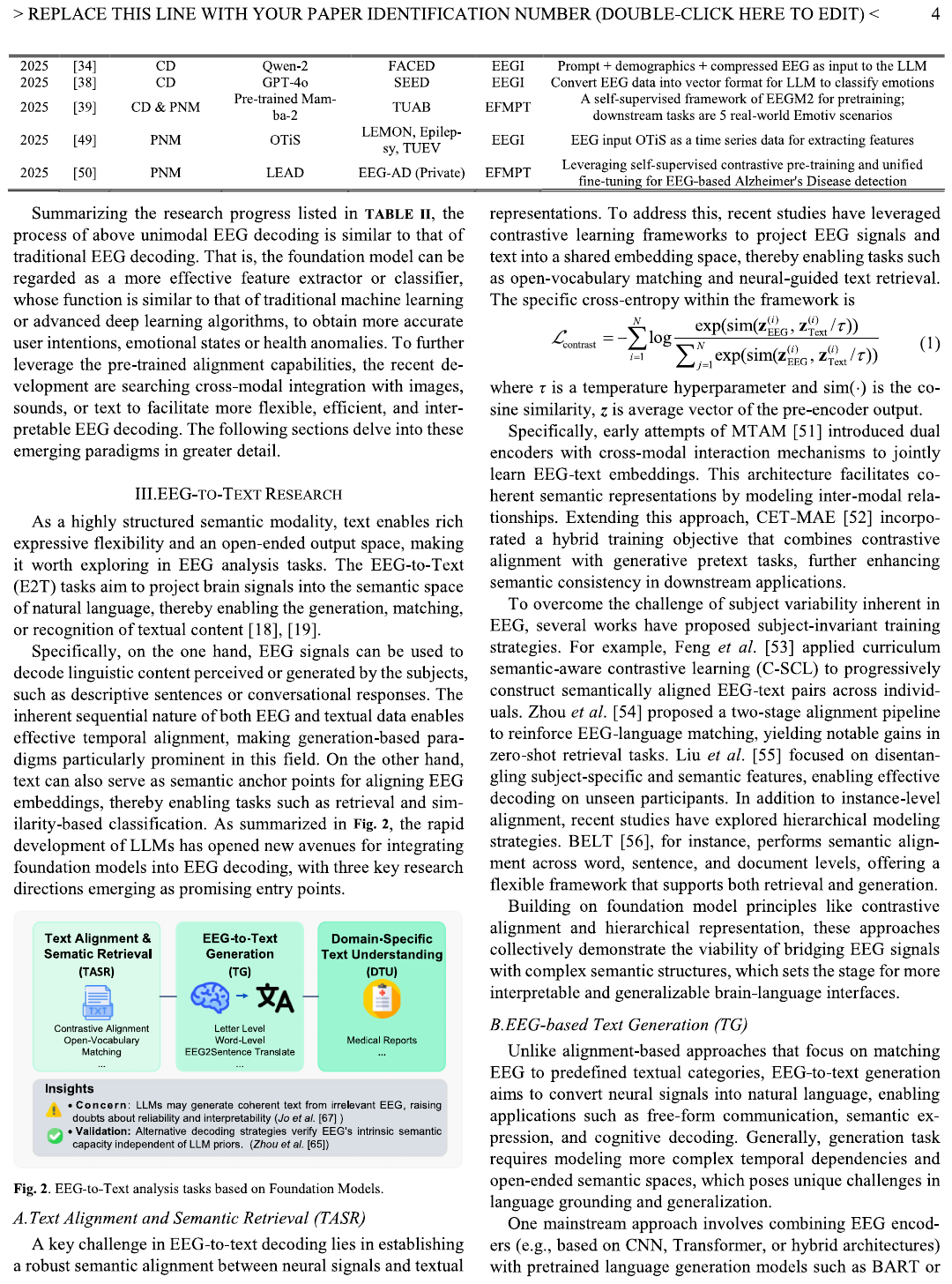}
\end{figure*}

Summarizing the research progress listed in \textbf{TABLE II}, the process of above unimodal EEG decoding is similar to that of traditional EEG decoding. That is, the foundation model can be regarded as a more effective feature extractor or classifier, whose function is similar to that of traditional machine learning or advanced deep learning algorithms, to obtain more accurate user intentions, emotional states or health anomalies. To further leverage the pre-trained alignment capabilities, the recent development are searching cross-modal integration with images, sounds, or text to facilitate more flexible, efficient, and interpretable EEG decoding. The following sections delve into these emerging paradigms in greater detail.

\section{EEG-to-Text Research}

As a highly structured semantic modality, text enables rich expressive flexibility and an open-ended output space, making it worth exploring in EEG analysis tasks. The EEG-to-Text (E2T) tasks aim to project brain signals into the semantic space of natural language, thereby enabling the generation, matching, or recognition of textual content \cite{18,19}.

\begin{figure}[h]
\centering
\includegraphics[width=8.8cm,
 trim=5mm 0mm 0mm 0mm, 
 clip]{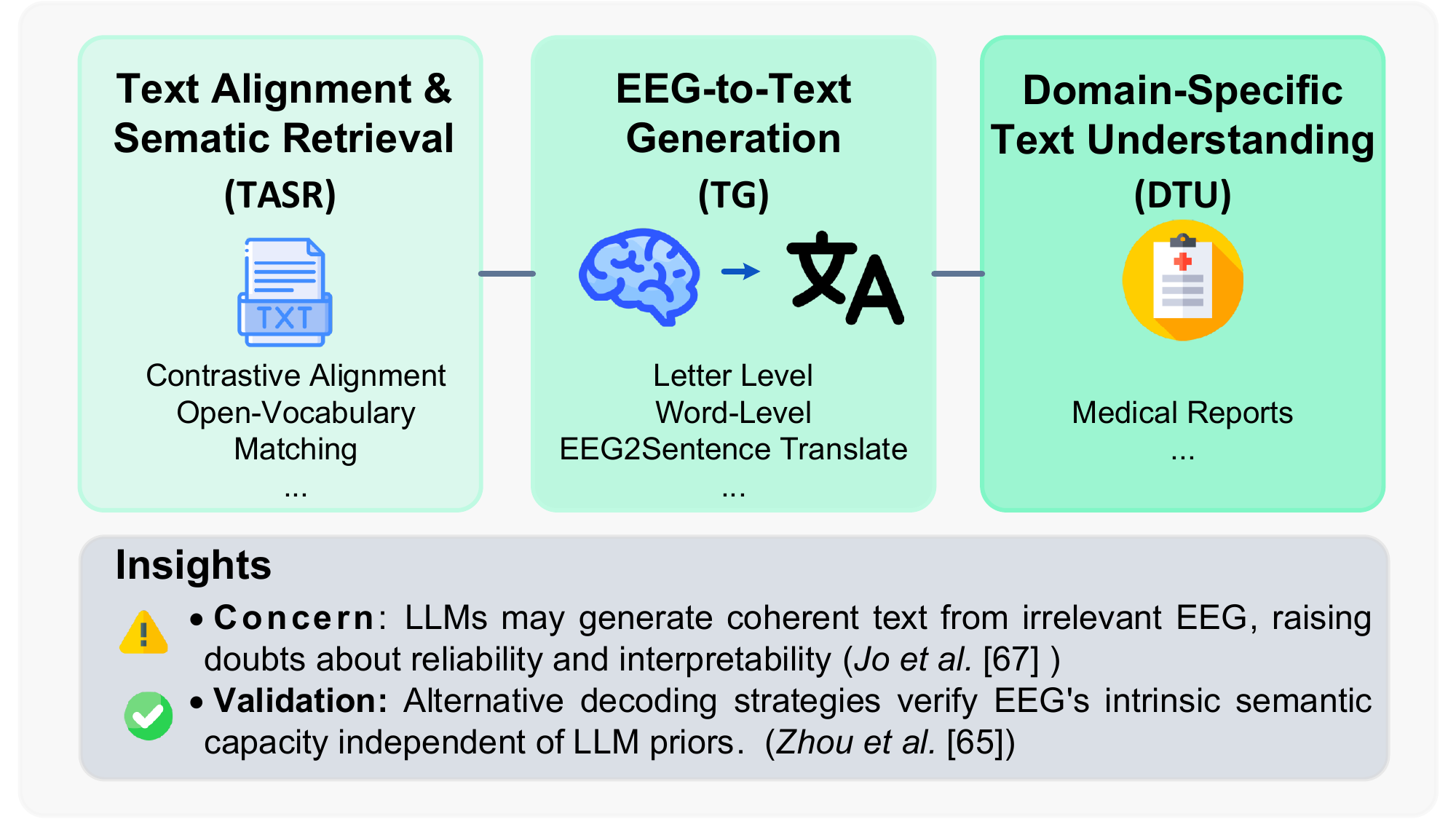}
\captionsetup{font=footnotesize}
\caption{EEG-to-Text analysis tasks based on Foundation Models.}
\label{fig 2}
\end{figure}

Specifically, on the one hand, EEG signals can be used to decode linguistic content perceived or generated by the subjects, such as descriptive sentences or conversational responses. The inherent sequential nature of both EEG and textual data enables effective temporal alignment, making generation-based paradigms particularly prominent in this field. On the other hand, text can also serve as semantic anchor points for aligning EEG embeddings, thereby enabling tasks such as retrieval and similarity-based classification. As summarized in Fig. 2, the rapid development of LLMs has opened new avenues for integrating foundation models into EEG decoding, with three key research directions emerging as promising entry points.

\subsection{Text Alignment and Semantic Retrieval (TASR)}

A key challenge in EEG-to-text decoding lies in establishing a robust semantic alignment between neural signals and textual representations. To address this, recent studies have leveraged contrastive learning frameworks to project EEG signals and text into a shared embedding space, thereby enabling tasks such as open-vocabulary matching and neural-guided text retrieval. The specific cross-entropy within the framework is

{\fontsize{8pt}{9.6pt}\selectfont
\[
\mathcal{L}_{\text{contrast}} = - \sum_{i=1}^N \log \frac{\exp\left(\mathrm{sim}\left(\mathbf{z}^{(i)}_{\mathrm{EEG}}, \mathbf{z}^{(i)}_{\mathrm{Text}} / \tau \right)\right)}{\sum_{j=1}^N \exp\left(\mathrm{sim}\left(\mathbf{z}^{(i)}_{\mathrm{EEG}}, \mathbf{z}^{(j)}_{\mathrm{Text}} / \tau \right)\right)}
\]
}

where $\tau$ is a temperature hyperparameter and \(f_{\mathrm{sim}}(\cdot)\) is the cosine similarity, z is average vector of the pre-encoder out-put.
Specifically, early attempts of MTAM \cite{51} introduced dual encoders with cross-modal interaction mechanisms to jointly learn EEG-text embeddings. This architecture facilitates coherent semantic representations by modeling inter-modal relationships. Extending this approach, CET-MAE \cite{52} incorporated a hybrid training objective that combines contrastive alignment with generative pretext tasks, further enhancing semantic consistency in downstream applications.

To overcome the challenge of subject variability inherent in EEG, several works have proposed subject-invariant training strategies. For example, Feng et al. \cite{53} applied curriculum semantic-aware contrastive learning (C-SCL) to progressively construct semantically aligned EEG-text pairs across individuals. Zhou et al. \cite{54} proposed a two-stage alignment pipeline to reinforce EEG-language matching, yielding notable gains in zero-shot retrieval tasks. Liu et al. \cite{55} focused on disentangling subject-specific and semantic features, enabling effective decoding on unseen participants. In addition to instance-level alignment, recent studies have explored hierarchical modeling strategies. BELT \cite{56}, for instance, performs semantic alignment across word, sentence, and document levels, offering a flexible framework that supports both retrieval and generation.

Building on foundation model principles like contrastive alignment and hierarchical representation, these approaches collectively demonstrate the viability of bridging EEG signals with complex semantic structures, which sets the stage for more interpretable and generalizable brain-language interfaces.

\subsection{EEG-based Text Generation (TG)}
Unlike alignment-based approaches that focus on matching EEG to predefined textual categories, EEG-to-text generation aims to convert neural signals into natural language, enabling applications such as free-form communication, semantic expression, and cognitive decoding. Generally, generation task requires modeling more complex temporal dependencies and open-ended semantic spaces, which poses unique challenges in language grounding and generalization.
One mainstream approach involves combining EEG encoders (e.g., based on CNN, Transformer, or hybrid architectures) with pretrained language generation models such as BART or  

GPT. For example, BART-based frameworks \cite{57,58} utilize encoded EEG vectors as input to the decoder, enabling fluent and stylistically coherent text generation. The Brain2Qwerty system \cite{59,60} integrates a deep EEG encoder with GPT-2 to achieve letter-level generation from EEG/MEG data, demonstrating robust generalization. GPT-4 has also been employed as a post-processing module to refine sentence fluency and semantic correctness \cite{61}. 

Some studies introduce cross-modal alignment mechanisms \cite{52,53,62} to further enhance consistency between EEG representations and textual semantic spaces, significantly improving generation quality. Beyond alignment-based models, some approaches adopt self-supervised learning or discrete encoding strategies. For example, EEG2TEXT \cite{63} employs multi-view Transformers and pretraining objectives to enhance EEG semantic representation without requiring textual supervision. DeWave \cite{64} introduces discrete latent representations to bridge EEG signals and token-level language embeddings. BELT-2 \cite{65} proposes a unified multitask framework leveraging byte pair encoding (BPE) and a shared decoder for text generation, translation, and affective recognition, highlighting the transferability of EEG-derived semantics across tasks.

Collectively, as we see from \textbf{TABLE III}, these mentioned EEG-to-text research leverages BART, BERT, and hybrid architectures (e.g., Conformer) with contrastive learning for semantic alignment. Trends show a shift toward generalization (zero-shot learning) and clinical applications (TUAB dataset). Future work should explore multimodal integration, interpretability, and low-resource adaptation to enhance scalability and practical utility across diverse domains.

\begin{figure*}[t]
  \centering
  \includegraphics[
    pagebox=cropbox, 
    clip,
    trim = 16mm 205mm 16mm 30mm, 
    width=\textwidth 
  ]{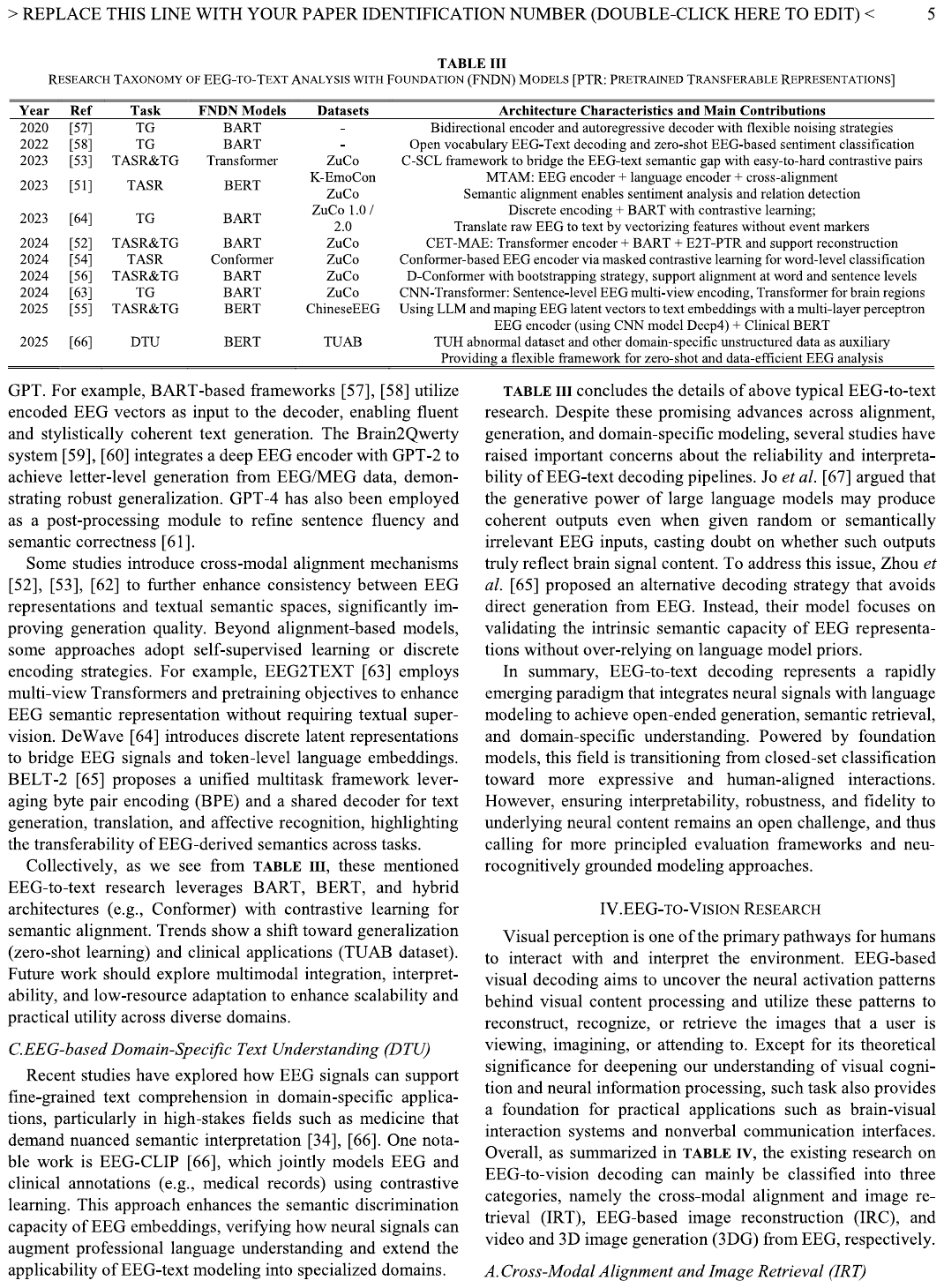}
\end{figure*}

\subsection{EEG-based Domain-Specific Text Understanding (DTU)}

Recent studies have explored how EEG signals can support fine-grained text comprehension in domain-specific applications, particularly in high-stakes fields such as medicine that demand nuanced semantic interpretation \cite{34,66}. One notable work is EEG-CLIP \cite{66}, which jointly models EEG and clinical annotations (e.g., medical records) using contrastive learning. This approach enhances the semantic discrimination capacity of EEG embeddings, verifying how neural signals can augment professional language understanding and extend the applicability of EEG-text modeling into specialized domains.

\textbf{TABLE III} concludes the details of above typical EEG-to-text research. Despite these promising advances across alignment, generation, and domain-specific modeling, several studies have raised important concerns about the reliability and interpretability of EEG-text decoding pipelines. Jo et al. \cite{67} argued that the generative power of large language models may produce coherent outputs even when given random or semantically irrelevant EEG inputs, casting doubt on whether such outputs truly reflect brain signal content. To address this issue, Zhou et al. \cite{65} proposed an alternative decoding strategy that avoids direct generation from EEG. Instead, their model focuses on validating the intrinsic semantic capacity of EEG representations without over-relying on language model priors.

In summary, EEG-to-text decoding represents a rapidly emerging paradigm that integrates neural signals with language modeling to achieve open-ended generation, semantic retrieval, and domain-specific understanding. Powered by foundation models, this field is transitioning from closed-set classification toward more expressive and human-aligned interactions. However, ensuring interpretability, robustness, and fidelity to underlying neural content remains an open challenge, and thus calling for more principled evaluation frameworks and neurocognitively grounded modeling approaches.

\section{EEG-to-Vision Research}
Visual perception is one of the primary pathways for humans to interact with and interpret the environment. EEG-based visual decoding aims to uncover the neural activation patterns behind visual content processing and utilize these patterns to reconstruct, recognize, or retrieve the images that a user is viewing, imagining, or attending to. Except for its theoretical significance for deepening our understanding of visual cognition and neural information processing, such task also provides a foundation for practical applications such as brain-visual interaction systems and nonverbal communication interfaces. Overall, as summarized in \textbf{TABLE IV}, the existing research on EEG-to-vision decoding can mainly be classified into three categories, namely the cross-modal alignment and image retrieval (IRT), EEG-based image reconstruction (IRC), and video and 3D image generation (3DG) from EEG, respectively.
\subsection{Cross-Modal Alignment and Image Retrieval (IRT)}

\begin{figure}[t]
\centering
\includegraphics[width=8.8cm,
 trim=20mm 0mm 10mm 0mm, 
 clip]{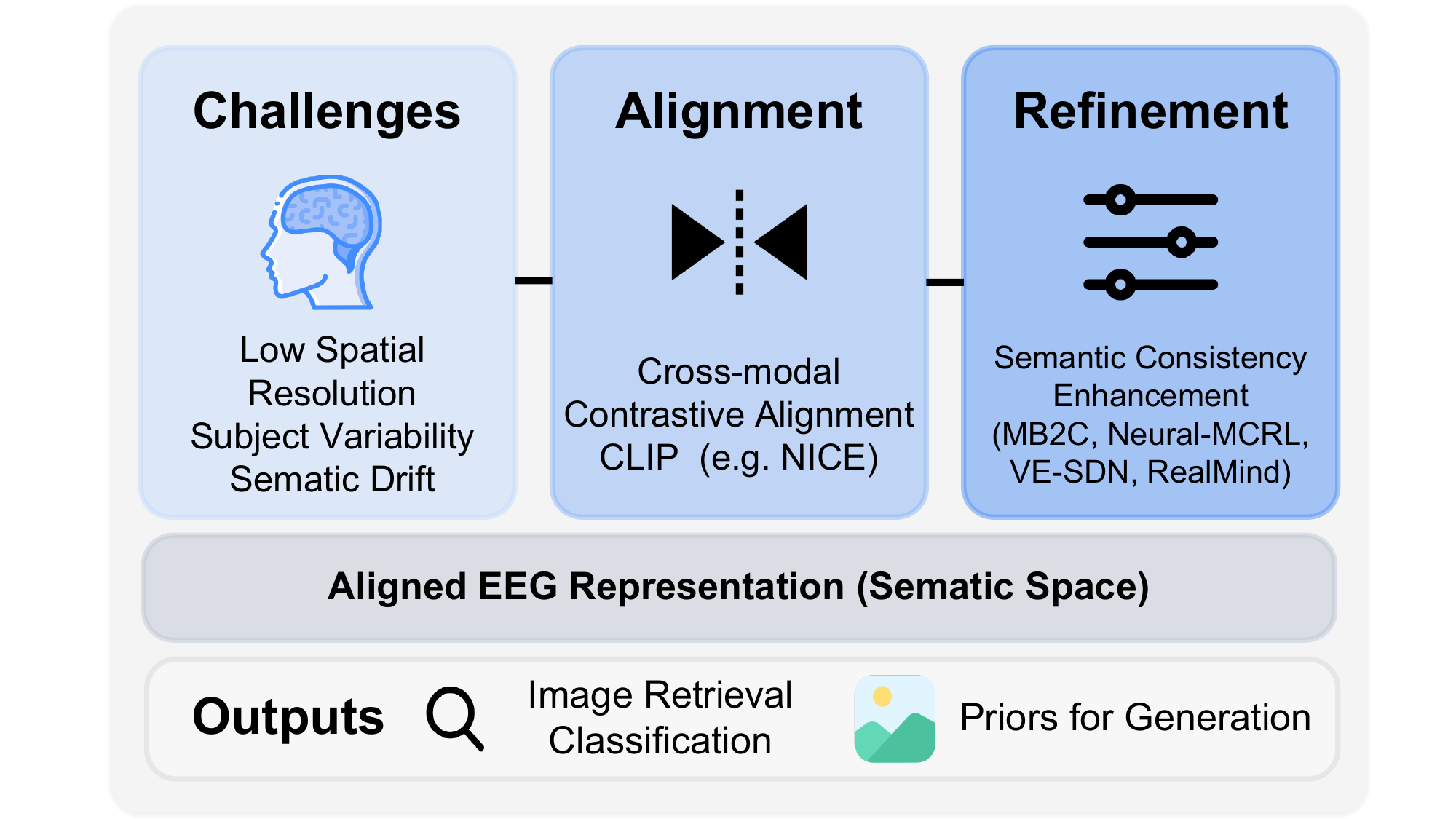}
\captionsetup{font=footnotesize}
\caption{The challenge and research idea of cross-modal alignment and image retrieval tasks for EEG based on Foundation Models.}
\label{fig 3}
\end{figure}

\begin{figure*}[b]
  \centering
  \includegraphics[
    pagebox=cropbox, 
    clip,
    trim = 16mm 28mm 16mm 172mm, 
    width=\textwidth 
  ]{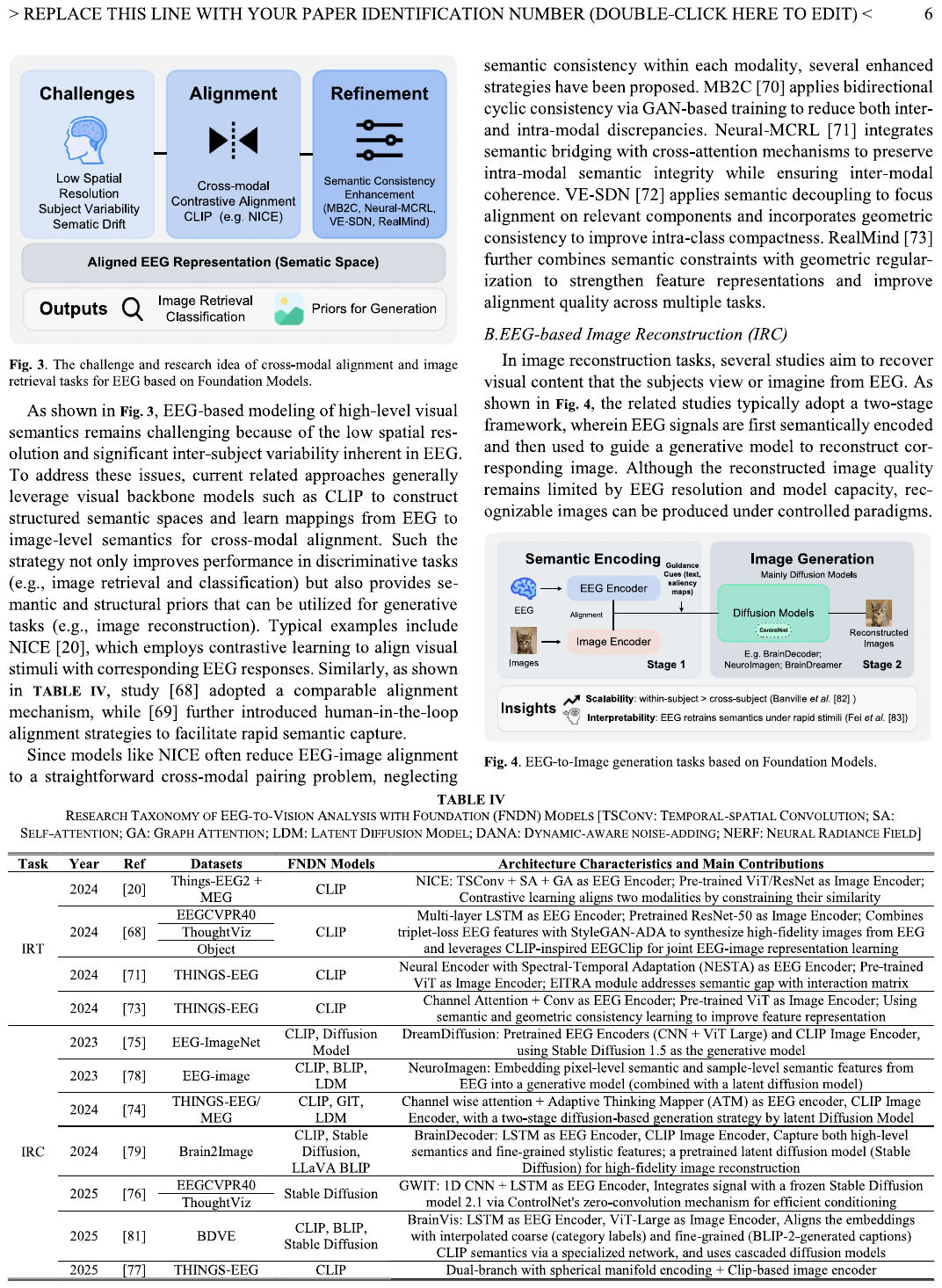}
\end{figure*}

As shown in \textbf{Fig. 3}, EEG-based modeling of high-level visual semantics remains challenging because of the low spatial resolution and significant inter-subject variability inherent in EEG. To address these issues, current related approaches generally leverage visual backbone models such as CLIP to construct structured semantic spaces and learn mappings from EEG to image-level semantics for cross-modal alignment. Such the strategy not only improves performance in discriminative tasks (e.g., image retrieval and classification) but also provides semantic and structural priors that can be utilized for generative tasks (e.g., image reconstruction). Typical examples include NICE \cite{20}, which employs contrastive learning to align visual stimuli with corresponding EEG responses. Similarly, as shown in TABLE IV, study \cite{68} adopted a comparable alignment mechanism, while \cite{69} further introduced human-in-the-loop alignment strategies to facilitate rapid semantic capture.

Since models like NICE often reduce EEG-image alignment to a straightforward cross-modal pairing problem, neglecting semantic consistency within each modality, several enhanced strategies have been proposed. MB2C \cite{70} applies bidirectional cyclic consistency via GAN-based training to reduce both inter- and intra-modal discrepancies. Neural-MCRL \cite{71} integrates semantic bridging with cross-attention mechanisms to preserve intra-modal semantic integrity while ensuring inter-modal coherence. VE-SDN \cite{72} applies semantic de-coupling to focus alignment on relevant components and in-corporates geometric consistency to improve intra-class compactness. RealMind \cite{73} further combines semantic constraints with geometric regularization to strengthen feature representations and improve alignment quality across multiple tasks.

\subsection{EEG-based Image Reconstrution (IRC)}
In image reconstruction tasks, several studies aim to recover visual content that the subjects view or imagine from EEG. As shown in \textbf{Fig. 4}, the related studies typically adopt a two-stage framework, wherein EEG signals are first semantically encoded and then used to guide a generative model to reconstruct corresponding image. Although the reconstructed image quality remains limited by EEG resolution and model capacity, recognizable images can be produced under controlled paradigms.

\begin{figure*}[t]
  \centering
  \includegraphics[
    pagebox=cropbox, 
    clip,
    trim = 16mm 245mm 16mm 30mm, 
    width=\textwidth 
  ]{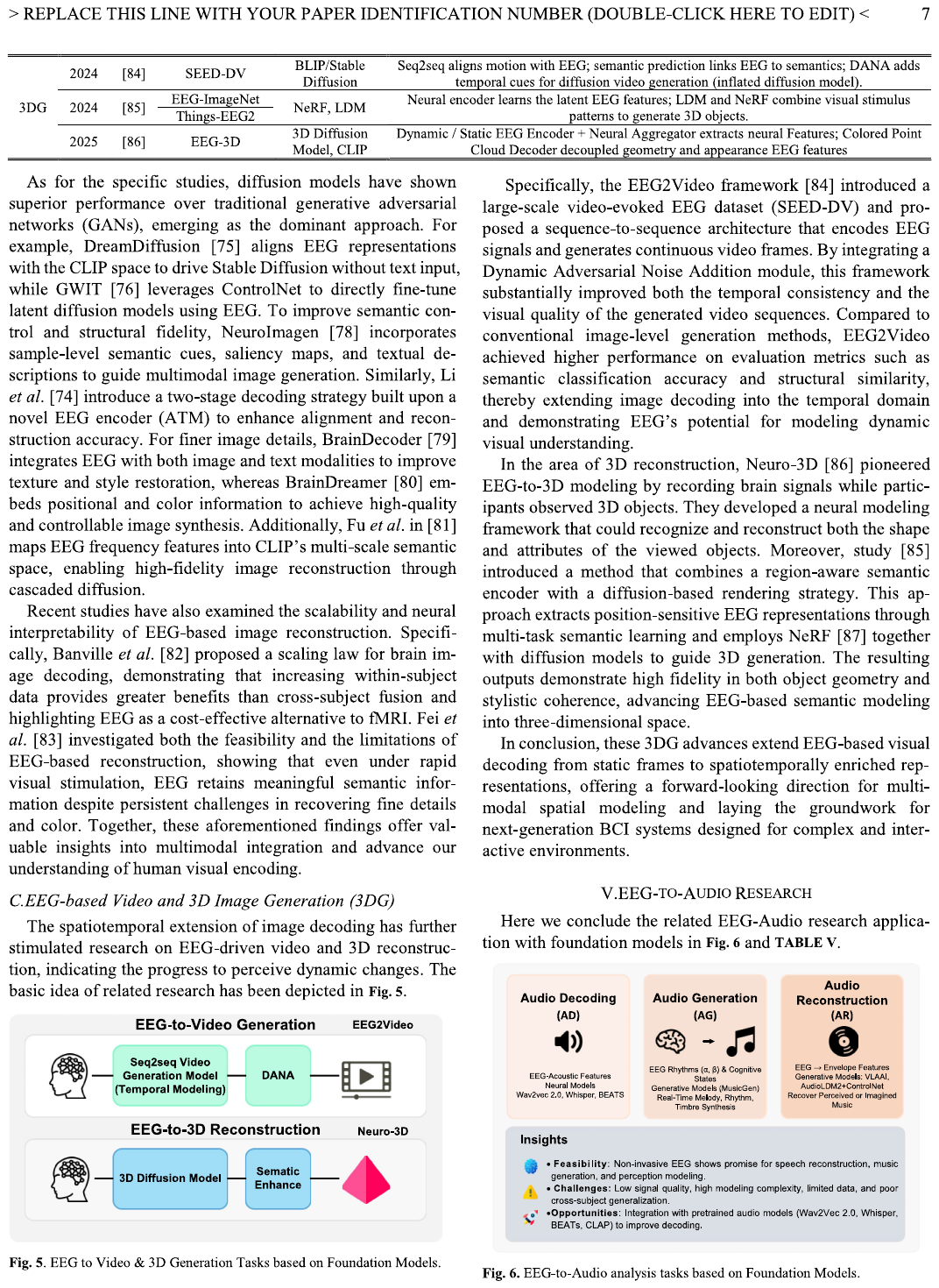}
\end{figure*}

As for the specific studies, diffusion models have shown superior performance over traditional generative adversarial networks (GANs), emerging as the dominant approach. For example, DreamDiffusion \cite{75} aligns EEG representations with the CLIP space to drive Stable Diffusion without text input, while GWIT \cite{76} leverages ControlNet to directly fine-tune latent diffusion models using EEG. To improve semantic control and structural fidelity, NeuroImagen \cite{78} incorporates sample-level semantic cues, saliency maps, and textual descriptions to guide multimodal image generation. Similarly, Li et al. \cite{74} introduce a two-stage decoding strategy built upon a novel EEG encoder (ATM) to enhance alignment and reconstruction accuracy. For finer image details, BrainDecoder \cite{79} integrates EEG with both image and text modalities to improve texture and style restoration, whereas BrainDreamer \cite{80} embeds positional and color information to achieve high-quality and controllable image synthesis. Additionally, Fu et al. in \cite{81} maps EEG frequency features into CLIP’s multi-scale semantic space, enabling high-fidelity image reconstruction through cascaded diffusion.

\begin{figure}[h]
\centering
\includegraphics[width=8.8cm,
 trim=0mm 30mm 0mm 0mm, 
 clip]{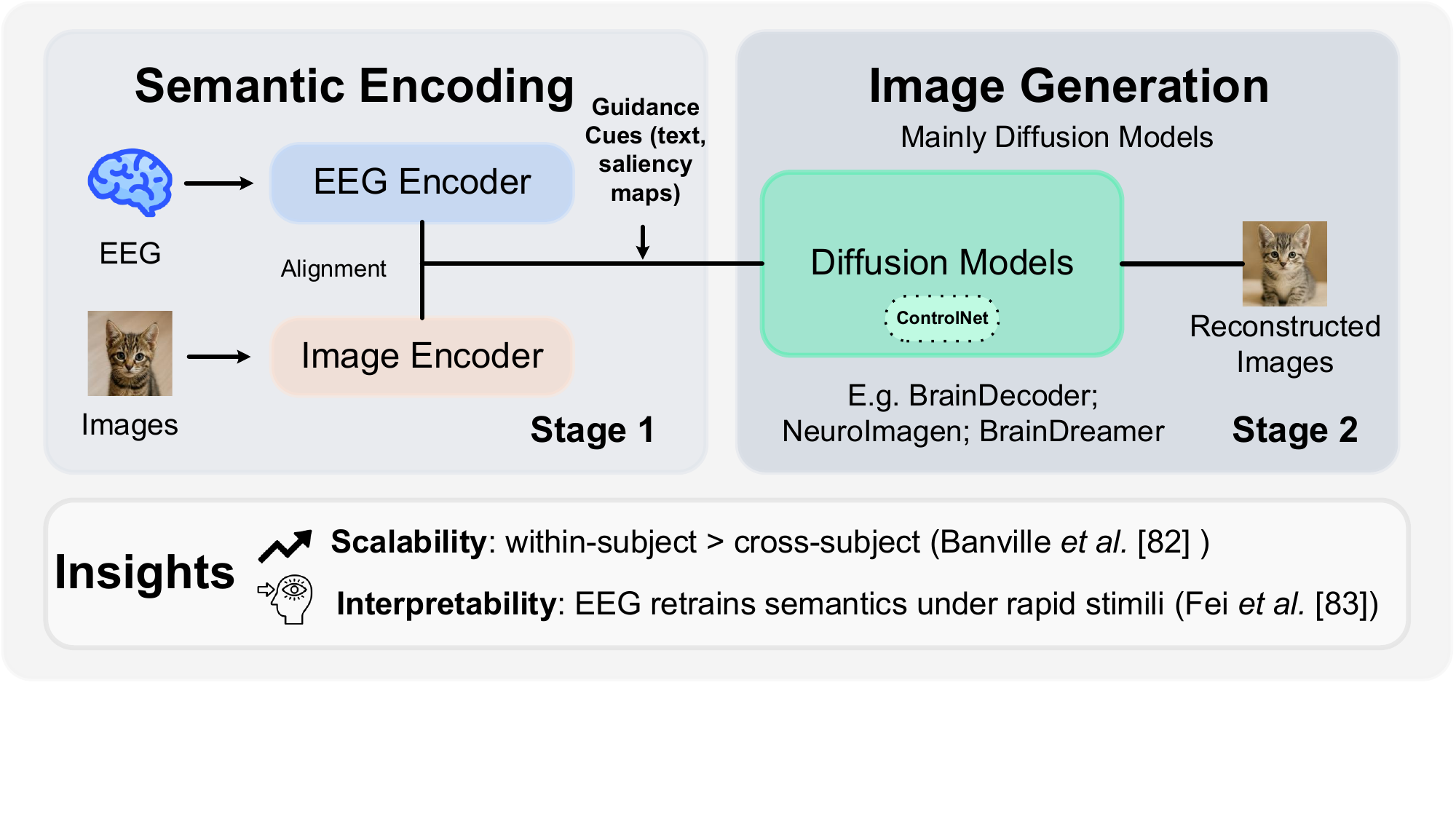}
\captionsetup{font=footnotesize}
\caption{EEG-to-Image generation tasks based on Foundation Models.}
\label{fig 4}
\end{figure}

Recent studies have also examined the scalability and neural interpretability of EEG-based image reconstruction. Specifically, Banville et al. \cite{82} proposed a scaling law for brain image decoding, demonstrating that increasing within-subject data provides greater benefits than cross-subject fusion and highlighting EEG as a cost-effective alternative to fMRI. Fei et al. \cite{83} investigated both the feasibility and the limitations of EEG-based reconstruction, showing that even under rapid visual stimulation, EEG retains meaningful semantic information despite persistent challenges in recovering fine details and color. Together, these aforementioned findings offer valuable insights into multimodal integration and advance our understanding of human visual encoding.

\subsection{EEG-based Video and 3D Image Generation (3DG)}
The spatiotemporal extension of image decoding has further stimulated research on EEG-driven video and 3D reconstruc-tion, indicating the progress to perceive dynamic changes. The basic idea of related research has been depicted in Fig. 5.

Specifically, the EEG2Video framework \cite{84} introduced a large-scale video-evoked EEG dataset (SEED-DV) and proposed a sequence-to-sequence architecture that encodes EEG signals and generates continuous video frames. By integrating a Dynamic Adversarial Noise Addition module, this framework substantially improved both the temporal consistency and the visual quality of the generated video sequences. Compared to conventional image-level generation methods, EEG2Video achieved higher performance on evaluation metrics such as semantic classification accuracy and structural similarity, thereby extending image decoding into the temporal domain and demonstrating EEG’s potential for modeling dynamic visual understanding.

\begin{figure}[h]
\centering
\includegraphics[width=8.8cm,
 trim=0mm 30mm 0mm 0mm, 
 clip]{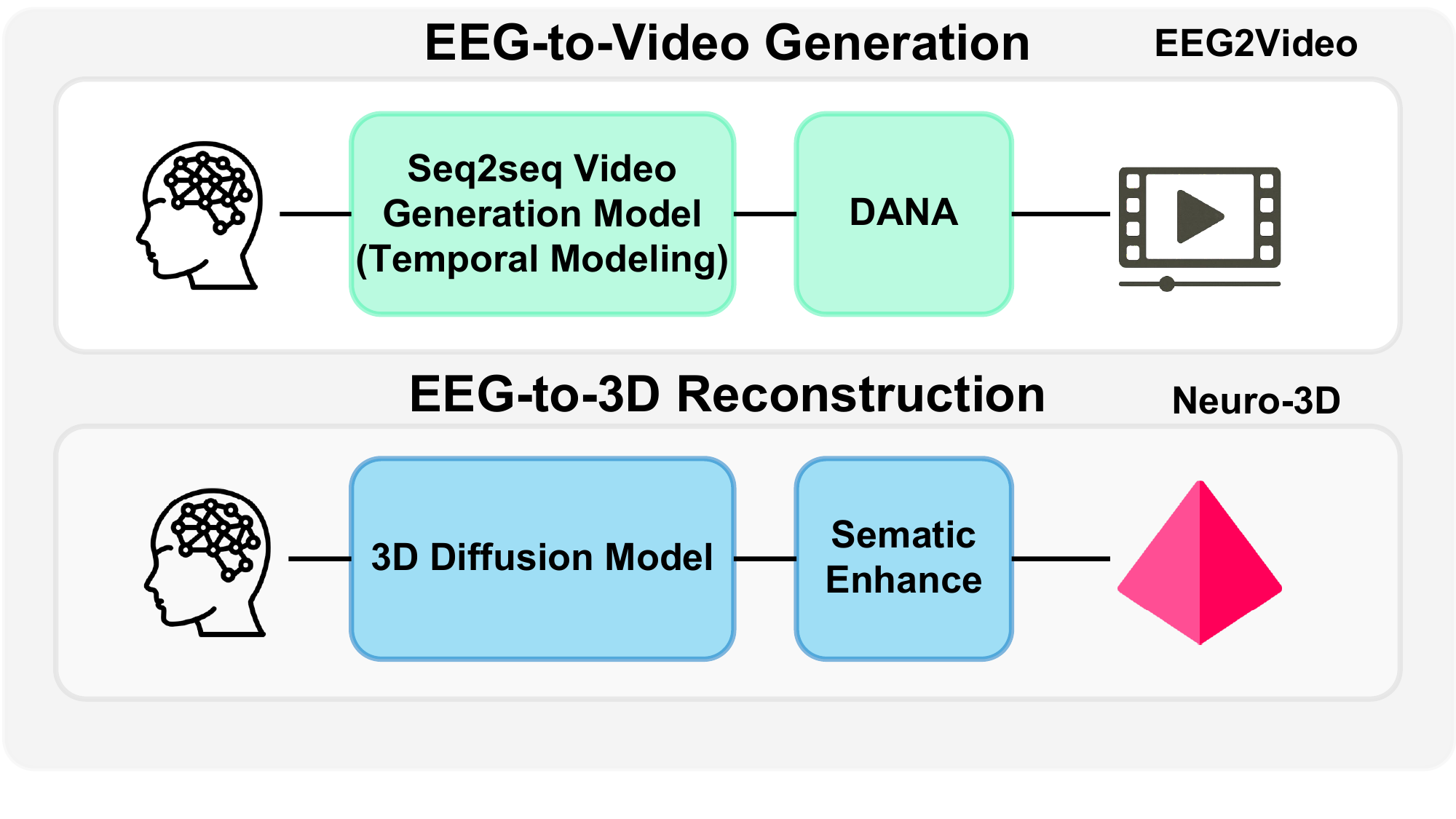}
\captionsetup{font=footnotesize}
\caption{EEG to Video \& 3D Generation Tasks based on Foundation Models.}
\label{fig 5}
\end{figure}

In the area of 3D reconstruction, Neuro-3D \cite{86} pioneered EEG-to-3D modeling by recording brain signals while participants observed 3D objects. They developed a neural modeling framework that could recognize and reconstruct both the shape and attributes of the viewed objects. Moreover, study \cite{85} introduced a method that combines a region-aware semantic encoder with a diffusion-based rendering strategy. This approach extracts position-sensitive EEG representations through multi-task semantic learning and employs NeRF \cite{87} together with diffusion models to guide 3D generation. The resulting outputs demonstrate high fidelity in both object geometry and stylistic coherence, advancing EEG-based semantic modeling into three-dimensional space.

In conclusion, these 3DG advances extend EEG-based visual decoding from static frames to spatiotemporally enriched representations, offering a forward-looking direction for multimodal spatial modeling and laying the groundwork for next-generation BCI systems designed for complex and interactive environments.

\section{EEG-TO-AUDIO RESEARCH}
Here we conclude the related EEG-Audio research application with foundation models in \textbf{Fig. 6} and \textbf{TABLE V}.

\subsection{EEG-based Audio Decoding (AD)}

Traditional high-fidelity audio decoding studies has relied more on invasive neural recordings such as electrocorticography (ECoG) and stereo-EEG (sEEG) \cite{88,89}, due to their superior spatial resolution and signal-to-noise ratio for obtaining detailed speech dynamics. With the continuous technique progress, especially the development of foundation models, recent findings have demonstrated the feasibility and growing potential of non-invasive EEG for audio decoding \cite{90}. Similar to EEG-Vision task, EEG-to-audio decoding focuses on identifying or reconstructing semantically relevant auditory information from EEG, encompassing speech and music that a user is listening to, imagining, or preparing to articulate. This research direction has been widely explored in automatic speech recognition and reconstruction \cite{91}, auditory attention detection \cite{92}, and music-related affective modeling \cite{93}. Typically, a Transformer-based autoregressive model of Generative Infinite-Vocabulary Transformer (GIVT) operating on compressed audio representations was proposed in \cite{93} to estimate musical surprisal (IC) directly from raw audio. Moreover, by leveraging a contrastive learning framework that aligns self-supervised speech representations through a shared convolutional architecture, study \cite{21} successfully decodes speech perception from non-invasive MEG and EEG recordings.

Beyond the above speech decoding, EEG-to-music decoding has also emerged as a complementary research direction that extends auditory decoding into creative and affective domains. Recent studies have explored music emotion recognition, leveraging multimodal fusion approaches with pretrained models such as CLAP and RoBERTa to jointly encode EEG, music audio, and lyrics, thereby enabling hierarchical cross-modal semantic alignment for affective modeling \cite{94}.

\begin{figure}[t]
\centering
\includegraphics[width=8.8cm,
 trim=0mm 0mm 50mm 0mm, 
 clip]{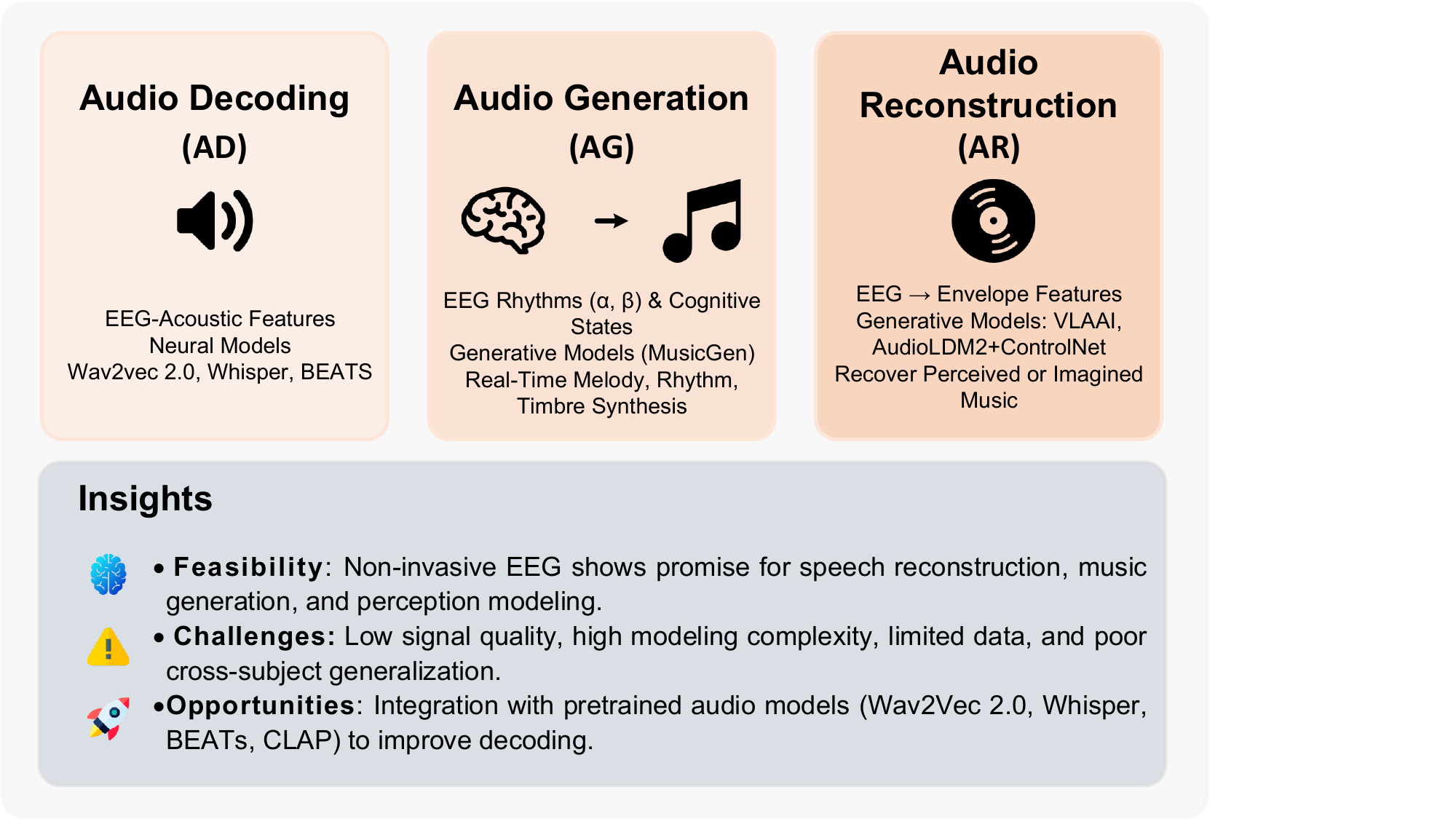}
\captionsetup{font=footnotesize}
\caption{EEG-to-Audio analysis tasks based on Foundation Models.}
\label{fig 6}
\end{figure}

\begin{figure*}[htbp]
  \centering
  \includegraphics[
    pagebox=cropbox, 
    clip,
    trim = 16mm 185mm 16mm 30mm, 
    width=\textwidth 
  ]{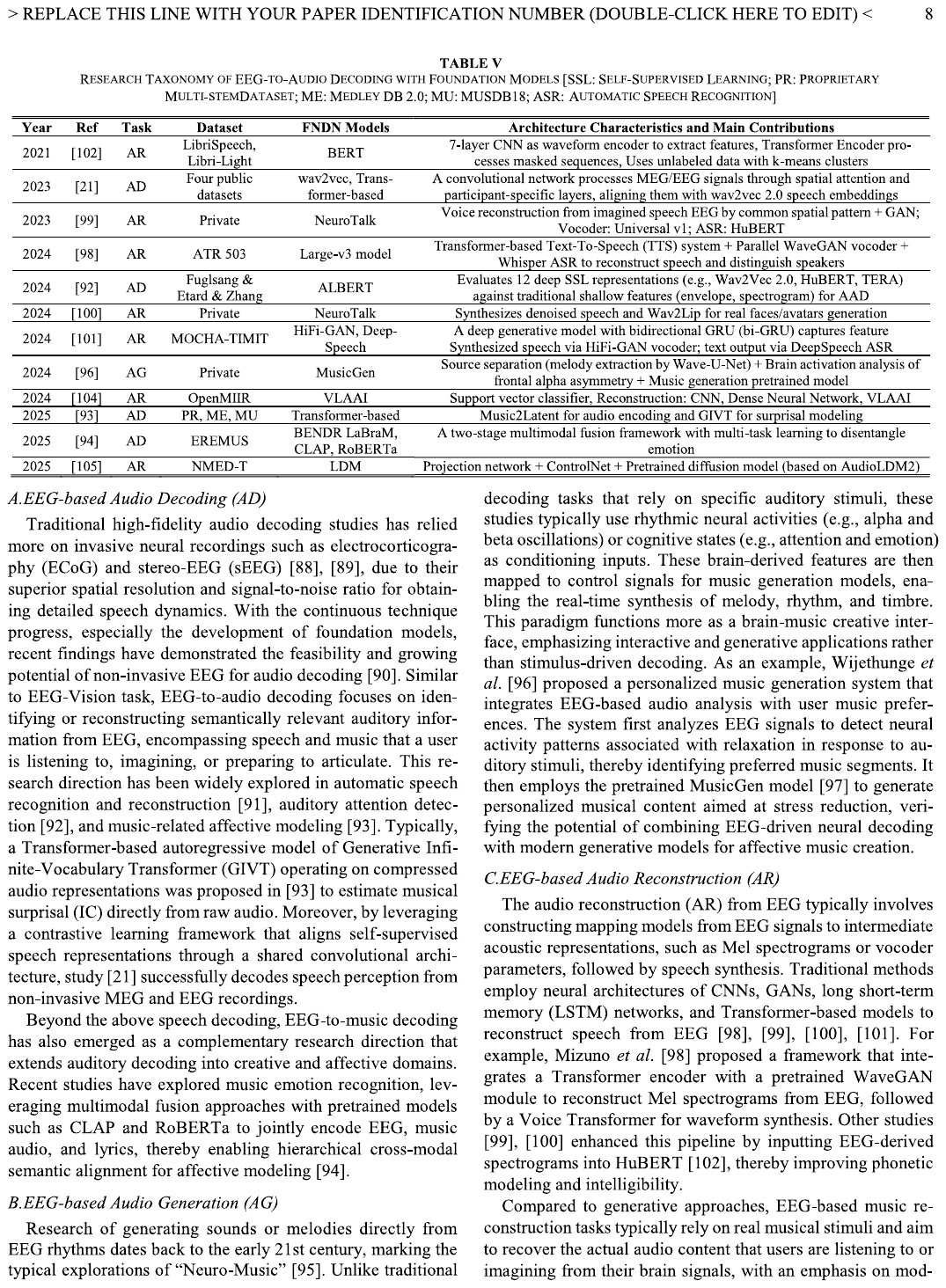}
\end{figure*}

\subsection{EEG-based Audio Generation (AG)}

Research of generating sounds or melodies directly from EEG rhythms dates back to the early 21st century, marking the typical explorations of “Neuro-Music” \cite{95}. Unlike traditional decoding tasks that rely on specific auditory stimuli, these studies typically use rhythmic neural activities (e.g., alpha and beta oscillations) or cognitive states (e.g., attention and emotion) as conditioning inputs. These brain-derived features are then mapped to control signals for music generation models, enabling the real-time synthesis of melody, rhythm, and timbre. This paradigm functions more as a brain-music creative interface, emphasizing interactive and generative applications rather than stimulus-driven decoding. As an example, Wijethunge et al. \cite{96} proposed a personalized music generation system that integrates EEG-based audio analysis with user music preferences. The system first analyzes EEG signals to detect neural activity patterns associated with relaxation in response to auditory stimuli, thereby identifying preferred music segments. It then employs the pretrained MusicGen model \cite{97} to generate personalized musical content aimed at stress reduction, verifying the potential of combining EEG-driven neural decoding with modern generative models for affective music creation.

\subsection{EEG-based Audio Reconstruction (AR)}
The audio reconstruction (AR) from EEG typically involves constructing mapping models from EEG signals to intermediate acoustic representations, such as Mel spectrograms or vocoder parameters, followed by speech synthesis. Traditional methods employ neural architectures of CNNs, GANs, long short-term memory (LSTM) networks, and Transformer-based models to reconstruct speech from EEG \cite{98,99,100,101}. For example, Mizuno et al. \cite{98} proposed a framework that integrates a Transformer encoder with a pretrained WaveGAN module to reconstruct Mel spectrograms from EEG, followed by a Voice Transformer for waveform synthesis. Other studies \cite{99,100} enhanced this pipeline by inputting EEG-derived spectrograms into HuBERT \cite{102}, thereby improving phonetic modeling and intelligibility.

Compared to generative approaches, EEG-based music re-construction tasks typically rely on real musical stimuli and aim to recover the actual audio content that users are listening to or imagining from their brain signals, with an emphasis on modeling the reproducibility of the auditory perception process. Early work by Bellier et al. \cite{103} employed intracranial EEG (iEEG) recordings from 29 participants who listened to Pink Floyd’s \textit{“Another Brick in the Wall, Part 1.”} Using a nonlinear multilayer perceptron (MLP) model, they successfully reconstructed recognizable audio fragments, thereby demonstrating the feasibility of reconstructing complex musical stimuli from short-term neural responses. Ankitha et al. \cite{104} proposed the VLAAI model, pretrained on large-scale audio datasets, to reconstruct the envelope features of musical stimuli from non-invasive EEG signals. This approach effectively addressed the limitations of traditional linear modeling in capturing complex temporal and spectral structures, thereby deepening our understanding of the mapping between brain activity and music perception. More recently, a diffusion-based generative paradigm has been applied to EEG-music reconstruction. Specifically, a framework combining AudioLDM2 and ControlNet \cite{100,105} uses EEG-derived features as conditional guidance to construct an end-to-end latent diffusion model without the need for manual channel selection or extensive preprocessing. Trained on the NMED-T natural music dataset, this method achieved significant improvements over multiple baselines in both perceptual quality and structural consistency of reconstructed audio.

Overall, as concluded in \textbf{TABLE V}, EEG-to-audio research are increasingly adopting transformer-based SSL models (e.g., wav2vec, LDM) and generative architectures (GANs, diffusion models) for speech/music reconstruction and attention decoding. The research trends include multimodal fusion (e.g., ControlNet) and emotion-aware synthesis, but reliance on private datasets poses reproducibility challenges. Future work should prioritize open benchmarks, standardized evaluation, and integration of EEG with other modalities (e.g., visual cues) to advance real-world applications like neuroprosthetics and affective computing.

\section{MULTI-MODAL EEG ANALYSIS TASKS}

Recently, multimodal EEG modeling has emerged as a hot focus in foundational research and has been widely applied to emotion recognition, semantic generation, and cognitive state modeling \cite{106,107,108}. Specifically, EEG-based multi-modal fusion tasks aim to integrate two or more modalities into a unified computational framework for collaborative modeling. Such the integration enables joint inference and multi-tasks learning. Since the relevant research leverages diverse information sources to improve model generalization in complex cognitive contexts, it places stronger emphasis on semantic complementarity, structural alignment, and cross-modal consistency. As summarized in \textbf{Fig. 7} and TABLE VI, according to the role of EEG signals, this section reviews four representative categories of related multimodal fusion tasks, highlighting their frameworks, key application, and used datasets.

\subsection{Multimodal Perception}
Due to the inherent difference between the foundation model purpose and the EEG task, introducing a new signal other than EEG during pre-training or use for the analysis (dual-modal approach) has become the basic idea of the field research.

\begin{figure}[t]
\centering
\includegraphics[width=8.8cm,
 trim=0mm 40mm 0mm 0mm, 
 clip]{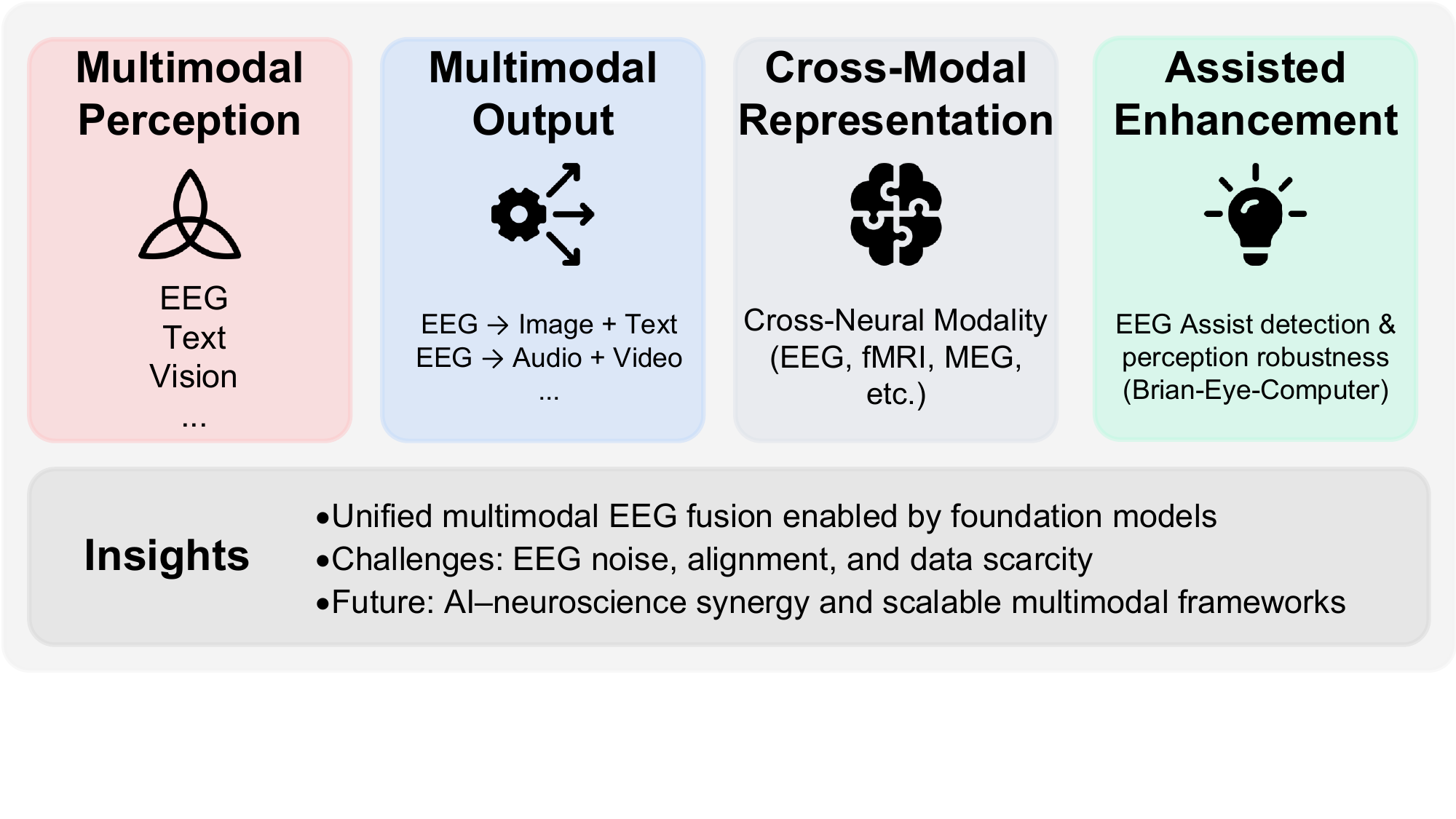}
\captionsetup{font=footnotesize}
\caption{Multi-modal EEG analysis tasks based on Foundation Models.}
\label{fig 7}
\end{figure}

Introducing a third or more modalities, such as text, audio, or physiological signals, on top of conventional EEG-language or EEG-vision bimodal framework can further enhance contextual understanding and semantic representation. For instance, CognitionCapturer of \cite{109} has integrated textual descriptions as semantic guidance supplements into the image reconstruction task, significantly improving scene completeness and reconstruction accuracy. Indeed, in the domain of music emotion recognition \cite{94}, we can also regard it combines music audio with corresponding modality of lyrics to facilitate hierarchical modeling of emotional dynamics. A reversed strategy has also been attempted. As an illustration, Thought2Text \cite{110} incorporates auxiliary image inputs during training to guide semantic alignment with EEG but relies solely on EEG during inference for text generation. Such the design underscores EEG’s autonomous capacity to express subjective thought while broadening the scope of multimodal fusion.

Moreover, the third modality can also serve as a supervisory signal to enforce semantic consistency across modalities, mitigating issues such as drift or misalignment between EEG and visual or linguistic representations (e.g., semantic bridging combined with cross-attention mechanisms in \cite{71} can simultaneously preserve intra-modal integrity and ensure inter-modal coherence). In terms of robustness, the third modality provides redundancy and complementary information when one modality suffers from noise or signal loss. The Milmer framework \cite{111}, as an illustration, integrates visual and physiological signals with a fine-tuned Swin Transformer, yielding substantial gains in emotion recognition accuracy and noise resilience, and demonstrating the practical benefits of tri-modal fusion.

\subsection{Multimodal Output}

This category of tasks employs EEG signals as the primary input to simultaneously generate multiple output modalities, such as images and text, with the aim of reconstructing a subject’s internal perceptions, imagined content, or semantic experiences. Compared to general unimodal generation tasks, this task highlights EEG’s potential to encode complex semantic structures and investigates how brain activity can be translated into multimodal representations. By doing so, it provides an indirect pathway to model cognitive states and mental experiences. Noting that although EEG remains the central input, several methods also integrate auxiliary signals, such as task descriptions or prompt labels, to improve controllability and enhance semantic coherence during generation.

\begin{figure*}[htbp]
  \centering
  \includegraphics[
    pagebox=cropbox, 
    clip,
    trim = 16mm 125mm 16mm 30mm, 
    width=\textwidth 
  ]{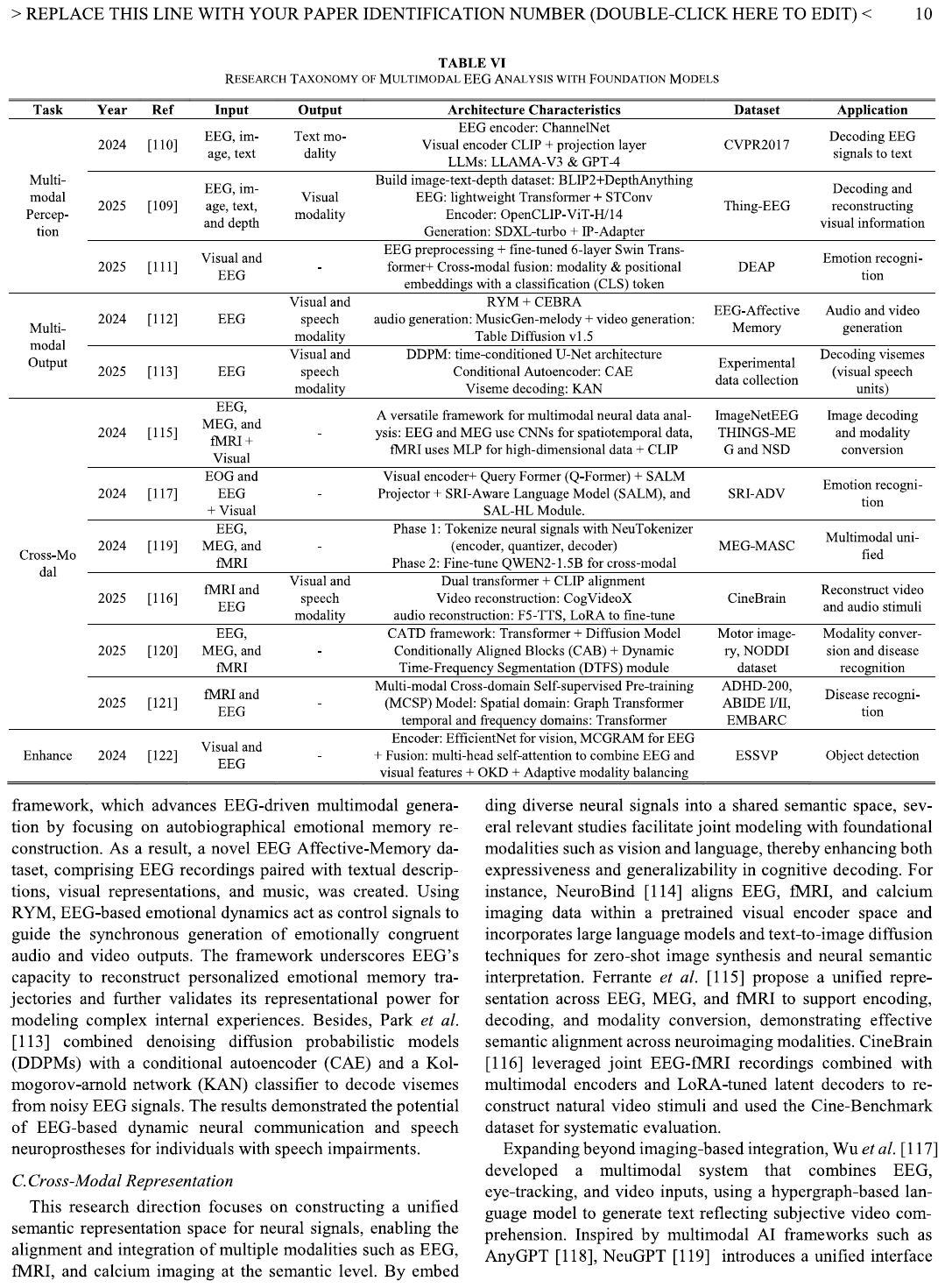}
\end{figure*}

Kwon et al. \cite{112} proposed the Recall Your Memory (RYM) framework, which advances EEG-driven multimodal generation by focusing on autobiographical emotional memory reconstruction. As a result, a novel EEG Affective-Memory dataset, comprising EEG recordings paired with textual descriptions, visual representations, and music, was created. Using RYM, EEG-based emotional dynamics act as control signals to guide the synchronous generation of emotionally congruent audio and video outputs. The framework under-scores EEG’s capacity to reconstruct personalized emotional memory trajectories and further validates its representational power for modeling complex internal experiences. Besides, Park et al. \cite{113} combined denoising diffusion probabilistic models (DDPMs) with a conditional autoencoder (CAE) and a Kolmogorov-arnold network (KAN) classifier to decode visemes from noisy EEG signals. The results demonstrated the potential of EEG-based dynamic neural communication and speech neuro-prostheses for individuals with speech impairments.

\subsection{Cross-Modal Representation}

This research direction focuses on constructing a unified semantic representation space for neural signals, enabling the alignment and integration of multiple modalities such as EEG, fMRI, and calcium imaging at the semantic level. By embedding diverse neural signals into a shared semantic space, several relevant studies facilitate joint modeling with foundational modalities such as vision and language, thereby enhancing both expressiveness and generalizability in cognitive decoding. For instance, NeuroBind \cite{114} aligns EEG, fMRI, and calcium imaging data within a pretrained visual encoder space and incorporates large language models and text-to-image diffusion techniques for zero-shot image synthesis and neural semantic interpretation. Ferrante et al. \cite{115} propose a unified representation across EEG, MEG, and fMRI to support encoding, decoding, and modality conversion, demonstrating effective semantic alignment across neuroimaging modalities. CineBrain \cite{116} leveraged joint EEG-fMRI recordings combined with multimodal encoders and LoRA-tuned latent decoders to reconstruct natural video stimuli and used the Cine-Benchmark dataset for systematic evaluation.

Expanding beyond imaging-based integration, Wu et al. \cite{117} developed a multimodal system that combines EEG, eye-tracking, and video inputs, using a hypergraph-based language model to generate text reflecting subjective video comprehension. Inspired by multimodal AI frameworks such as AnyGPT \cite{118}, NeuGPT \cite{119}  introduces a unified interface capable of mapping various neural signals to language and speech. Similarly, the Condition-Aligned Temporal Diffusion (CATD) framework \cite{120} synthesizes fMRI-detected Blood Oxygen Level Dependent (BOLD) signals from EEG using a Diffusion Transformer for cross-modal neural representation, while MCSP \cite{121} integrates EEG and fMRI across spatial, temporal, and frequency domains to advance multimodal neuroimaging analysis.

\subsection{Assisted Enhancement}

In certain studies, EEG signals are not utilized as the primary information source for semantic modeling or generation. Instead, they function as an auxiliary modality within multi-modal systems to help collaborative perception and decision-making. Such tasks typically focus on domains centered on images, audio, or other external modalities, while EEG contributes complementary signals that encode implicit in-formation related to individual cognitive states, attentional mechanisms, or neural responses. By incorporating these cues, models can achieve enhanced discriminative capability in complex scenarios, illustrating a strong intersection between cognitive modeling and engineering-driven augmentation.

As shown in the recently proposed Brain-Eye-Computer framework \cite{122}, EEG and eye-tracking data are integrated as auxiliary modalities within an image detection model to improve accuracy and robustness in weak target detection for remote sensing imagery. The framework establishes a tri-modal data channel encompassing brain, eye, and image inputs, and employs an online knowledge distillation (OKD) mechanism to guide the image encoder toward human attention regions, thereby achieving a detection strategy aligned with perceptual pathways observed in human vision. Although such approaches do not directly generate semantic outputs from EEG, they effectively highlight the role of EEG in improving perception accuracy and reliability through cross-modal fusion.

In summary, as we concluded in \textbf{TABLE VI}, multimodal EEG analysis with foundation models have demonstrated a clear trend toward integrating diverse inputs (EEG, fMRI, text) using transformer-based architectures and cross-modal fusion. As a matter of fact, through the use of cross-modal alignment techniques, the research area with foundation models is transitioning toward unified frameworks that support perception, generation, and neurocognitive interpretation, while light-weight designs entailed in each architecture are designed to enable real-time applications. 

\section{Discussion}

With the entry of foundation models in recent years, the EEG analysis has fundamentally transformed from traditional signal processing and pattern recognition pipelines, signifying a paradigm shift in EEG tasks of single-modality signal decoding toward multimodal integration and high-level cognitive inference. This survey systematically review the study of applying foundation models to EEG analysis across multiple tasks, including unimodal EEG decoding, EEG-to-Text, EEG-to-Vision, EEG-to-Audio, and multimodal EEG fusion. 

By integrating the contents of \textbf{TABLE II, TABLE III, TABLE IV, TABLE V}, and \textbf{TABLE VI}, we can summarize that a typical paradigm of foundation model based EEG analysis comprises three core components: the EEG encoder, the cross-modal alignment module, and the generative or discriminative decoder. Specifically for the architectural design, the EEG encoder is responsible for mapping raw signals into robust and separable feature spaces. Early approaches relied on CNNs, which are highly effective in capturing local spatiotemporal patterns but less capable of modeling long-range dependencies and inter-channel interactions. Transformer-based architectures, as well as hybrid CNN-Transformer designs, have gradually become dominant due to their ability to balance global dependency modeling and local detail preservation through multi-head attention \cite{5}. 

The cross-modal alignment module serves as the semantic bridge between EEG and the target modality, directly influencing downstream performance. Current alignment strategies include static mapping approaches that project features into a shared embedding space with contrastive losses \cite{20,52,53,90}, as well as dynamic methods leveraging cross-attention or Q-Former \cite{117} architectures to explicitly model token-level interactions. The latter demonstrates superior adaptability in generation tasks requiring fine-grained semantic control. In the decoding and generation stage, the text-related tasks benefit from LLMs to improve fluency and coherence \cite{58}. Vision tasks often integrate semantic priors from CLIP \cite{74}, BLIP \cite{79,81}, or diffusion models \cite{76,84}, while in audio tasks, pre-trained backbones such as MusicGen \cite{96} and AudioLDM2 \cite{105} show notable advantages in emotional consistency and perceptual fidelity.

From a training methodology perspective, foundation model-based EEG analysis typically follows a multi-stage paradigm of “self- or unsupervised pretraining, cross-modal alignment, task-specific fine-tuning.” In the self-supervised stage, methods such as contrastive learning \cite{20}, masked reconstruction \cite{52}, and multi-view consistency \cite{63} are used to learn cross-subject robust representations from large-scale unlabeled EEG data. The cross-modal alignment stage leverages limited paired data to align semantic spaces \cite{112}, often incorporating curriculum learning \cite{26,53}, cycle con-sistency \cite{70}, or semantic consistency regularization \cite{71} to reduce modality gaps. In the task-specialization stage, re-searchers balance performance and computational cost by choosing between full fine-tuning and parameter-efficient adaptation techniques such as LoRA \cite{116}, adapters \cite{44}, or prompt tuning \cite{34}. The unification of representation spaces not only enhances task transferability but also lays the groundwork for multi-task learning \cite{46}, domain adaptation, and data augmentation \cite{122}, and ultimately mitigating challenges related to data scarcity and domain shifts.

\subsection{Challenges}

Despite the notable progress brought by foundation models in EEG decoding, several critical challenges remain unresolved. The first challenge lies in insufficient cross-subject generalization, which continues to hinder real-world deployment. Due to differences in brain anatomy, cognitive strategies, and recording conditions, EEG signals exhibit substantial variability across individuals \cite{55,82}. Even with powerful pre-training, the learned representations often overfit to specific subject distributions, leading to significant performance degradation when applied to unseen individuals. The nonstationarity within the subjects further exacerbates this problem, where session-to-session variability can be as pronounced as inter-subject differences, limiting the reliability of zero-shot or few-shot transfer in practical applications. Be-sides, referring to the current field progress, rigorous cross-paradigm benchmarks and standardized evaluation protocols remain lacking \cite{123}.

Another key challenge concerns the authenticity and interpretability of cross-modal alignment mechanisms. As EEG signals fundamentally differ from high-level modalities such as text or vision in both structure and semantics. Existing pipelines often rely on post-hoc contrastive alignment or feature mapping, which may inadequately capture the neuro-semantic correspondence needed for robust decoding or analysis. Furthermore, the growing use of generative modeling raises concerns about semantic authenticity. While EEG-driven text, image, and music generation produce compelling outputs, it is unclear whether these outputs truly reflect EEG-encoded content or are predominantly shaped by the strong priors embedded in pretrained language and vision models \cite{67}. This ambiguity underscores the need for interpretability frameworks and neurophysiological validation to ensure that generative outputs are grounded in genuine neural representations \cite{65}.

Moreover, there remains a lack of research on EEG generative and inverse modeling. Most current research focuses on decoding EEG into other modalities, while the reverse process of generating EEG from visual, linguistic, or auditory inputs remains underexplored. Indeed, advancing such inverse modeling could enable closed-loop neurofeedback, facilitate hypothesis-driven neuroscientific studies, and provide new tools that validate neurocognitive theories through synthetic EEG reconstruction \cite{124}. However, relevant research attempts with fundamental models have not yet been widely carried out.

\subsection{Future Direction}

Addressing these challenges will require a concerted effort in architectural design, training methodology, evaluation protocols, and application development. Architecturally, future EEG encoders should integrate multi-scale time-frequency analysis, dynamic channel selection, and brain-region priors to more effectively capture discriminative neural patterns while maintaining robustness across different subjects and sessions. For the cross-modal alignment, developing EEG-specific interaction mechanisms, such as biologically constrained attention layers, neuro-symbolic reasoning modules, and hierarchical semantic alignment frameworks, can reduce semantic drift and ensure that generated outputs remain closely tied to the underlying neural activity. Plus, incorporating neurophysiological priors into alignment objectives may also enhance interpretability. One should instigate more effective generative or discriminative decoder. Finally, lightweight yet adaptive module designs, without compromising accuracy, could further enable real-time deployment in wearable BCI systems \cite{125}.

In terms of the training methodology, combining large-scale self-supervised pretraining on heterogeneous EEG corpora with domain-specific regularization techniques, such as cross-modal consistency, temporal coherence, and subject-aware parameter modulation, can improve robustness and transferability. Federated and continual learning strategies may additionally facilitate adaptation to new subjects and recording environments without catastrophic forgetting.

As for the evaluation, unified cross-task and cross-modal benchmarks \cite{123} incorporating both authenticity metrics (e.g., correlation with ground-truth neural patterns) and interpretability measures (e.g., attention alignment with known neuroanatomical structures), are essential for reproducibility and fair comparison. Furthermore, longitudinal evaluation protocols could further assess stability over extended periods, which is critical for deployment in real-world scenarios.

Finally, in the application development, we envision the emergence of EEG Digital Twins \cite{126}, with the personalized neural models trained on individual EEG recordings to replicate subject-specific cognitive and physiological dynamics. Meanwhile, the development of generative EEG models could bridge forward and reverse modeling, enabling not only “EEG-to-X” analysis but “X-to-EEG” synthesis for neuro-scientific simulation. On these basis, the integration of EEG-based systems with foundation model architectures holds strong potential for precise cognitive modeling, adaptive BCIs, and personalized neurorehabilitation. 

\section{Conclusion}
The fusion of EEG decoding with foundation models has initiated a paradigm shift from signal-centric methods toward semantically rich, multimodal, and generative frameworks. This paper presents a modality-oriented review of foundation models applied to EEG analysis. By consolidating advances across unimodal EEG decoding, EEG-to-text, EEG-to-vision, EEG-to-audio, and multimodal integration, this survey underscores both the feasibility, transformative potential, challenges, and future directions of this emerging research direction.

\end{document}